\documentclass[a4paper,twoside]{article}

\usepackage{epsfig}
\usepackage{subcaption}
\usepackage{calc}
\usepackage{amssymb}
\usepackage{amstext}
\usepackage{amsmath}
\usepackage{amsthm}
\usepackage{multicol}
\usepackage{pslatex}
\usepackage{apalike}
\usepackage{algorithm2e}
\usepackage[bottom]{footmisc}
\usepackage{SCITEPRESS}
\usepackage{graphicx}
\usepackage{float}
\usepackage[section]{placeins}

\begin{document}


\title{The Fifth Graph Normal Form (5GNF): A Trait-Based Framework for Metadata Normalization 
in Property Graphs}

\author{
\authorname{Yahya Sa'd\sup{1}, Vojtěch Merunka\sup{1,2}, Renzo Angles\sup{3}}
\affiliation{
\sup{1}Department of Information Engineering, Faculty of Economics and Management,\\
Czech University of Life Sciences Prague, Prague, Czech Republic\\
\sup{2}Department of Software Engineering, Faculty of Nuclear Sciences and Engineering,\\
Czech Technical University in Prague, Prague, Czech Republic\\
\sup{3}Department of Computer Science, Faculty of Engineering,\\
Universidad de Talca, Talca, Chile
}
\email{sady@pef.czu.cz, merunka@pef.czu.cz, vojtech.merunka@fjfi.cvut.cz, rangles@utalca.cl}
}


\begin{center}
\small
Preprint. Accepted for publication in the Proceedings of the 21st International Conference on
Evaluation of Novel Approaches to Software Engineering (ENASE 2026).
The final published version will appear in the SciTePress Digital Library.
\end{center}


\abstract{Graph databases are increasingly used in software systems that depend on 
rich metadata, yet current modelling practices often duplicate metadata across 
nodes, leading to redundancy and inconsistent semantics. This paper introduces 
5GNF, a trait-based normalization framework that structures metadata as canonical 
Trait Nodes linked through HAS\_TRAIT relationships. We formalize trait 
dependencies (tFDs), present the TraitExtraction5GNF algorithm, and implement 
the approach in Neo4j. To assess its practical relevance, we apply 5GNF to the 
widely used Northwind real dataset, which contains substantial duplicated 
location and shipping metadata. Normalization externalizes all recurring 
metadata into reusable traits, removes thousands of redundant attribute 
instances, reduces schema complexity, and simplifies analytical queries.
 Performance remains competitive due to more selective metadata access paths. 
 The results demonstrate that 5GNF provides a reproducible, semantically precise, 
 and engineering-oriented approach to metadata modelling, aligning well 
 with ENASE's focus on conceptual modelling and software design methodologies.}


\maketitle

\section{\uppercase{Introduction}}

Graph databases are increasingly central to modern information systems that rely
on highly interconnected data, including knowledge graphs, recommendation
systems, and research-information management platforms~\cite{ref9,ref10,ref13}.
While the property-graph model provides expressive modelling capabilities,
large-scale deployments frequently exhibit redundancy, inconsistency, and limited
schema flexibility—problems that become particularly visible when metadata is
repeatedly embedded across heterogeneous nodes and relationships~\cite{ref11,ref12}.

Throughout this work, metadata normalization applies uniformly to both nodes and 
relationships, since
 property graphs support properties on edges as first-class elements.

In practice, metadata describing provenance, temporal validity, regulatory
context, or organizational scope is often duplicated across graph schemas,
leading to increased maintenance costs and governance complexity~\cite{ref20,ref21}.
Existing graph-modelling practices offer limited support for modular and reusable
metadata representation, resulting in tightly coupled schemas that are difficult
to evolve.

Normalization theory addressed analogous issues in relational systems,
culminating in the Fifth Normal Form (5NF)~\cite{ref1}. Extensions of normalization
to graphs include Frisendal’s Graph Normal Form (GNF)~\cite{ref2} and the first
four graph normal forms (1GNF–4GNF) proposed by Merunka et al.~\cite{ref3}.
Subsequent work on graph functional dependencies (gFDs) and graph uniqueness
constraints (gUCs) demonstrated how dependency-based normalization can reduce
redundancy at the data level~\cite{ref4,ref5}. However, these approaches remain
entity-centric and do not address the normalization of reusable metadata
structures.

This paper introduces the \emph{Fifth Graph Normal Form (5GNF)}, a trait-based
normalization framework that elevates normalization from the data layer to the
metadata layer. The central construct is the \emph{Trait Node}, an atomic and
reusable schema component connected to graph elements via explicit
\texttt{HAS\_TRAIT} relationships. Guided by the Design Science Research
Methodology (DSRM)~\cite{ref14} and principles of composition-based 
modelling~\cite{ref15,ref16,ref17},
the framework formalizes conditions for metadata normalization and defines a
systematic trait-extraction and schema-rewriting process.

An experimental evaluation on real-world datasets demonstrates that 5GNF improves
metadata reuse and enhances schema modularity, supporting more maintainable and
interoperable graph designs. We position 5GNF as the natural endpoint of the
graph-normalization hierarchy (1GNF–5GNF)~\cite{ref4,ref5}, extending normalization 
theory beyond node- and edge-level data 
to the metadata structures that dominate contemporary graph applications.

While representing annotations or metadata as separate nodes is a known modeling
idiom in graph databases, our contribution is not the idiom itself. We introduce
a graph normal form that normatively characterizes when metadata should be
externalized, provide a dependency-based formalization in the form of trait
functional dependencies (tFDs) with explicit correctness criteria, and evaluate
the approach as a repeatable normalization procedure with measurable effects on
metadata redundancy, reuse, and query cost.

This work continues a long-term research effort on normalization across data and
metadata layers, building on earlier results in object-oriented conceptual
modeling and extending them to dependency-driven normalization of property
graphs.

The remainder of this paper is structured as follows. Section~2 reviews related
work. Section~3 formalizes the 5GNF framework. Section~4 introduces the
trait-extraction algorithm. Section~5 presents an illustrative example.
Section~6 reports the experimental evaluation, followed by implications in
Section~7, limitations in Section~8, and conclusions in Section~9.

\section{\uppercase{Background and Related Work}}
\label{sec:background}

\subsection{Normalization in Relational Databases}

Normalization has been a foundational concept in relational database theory since the 
introduction of the relational model. Successive normal forms—culminating in the Fifth Normal
 Form (5NF)—established principled mechanisms for reducing redundancy, preserving 
 dependencies, and ensuring lossless decomposition through functional, multivalued, and 
 join dependencies. These principles continue to inform modern schema-design practices.

\subsection{From Relational to Graph Normalization}

As graph databases gained adoption, normalization concepts were extended beyond tabular 
structures.  
Frisendal’s Graph Normal Form (GNF) introduced early normalization principles for graphs,
 emphasising atomicity and semantic clarity.  
Merunka et al.\ later formalised the first four Graph Normal Forms (1GNF–4GNF), addressing
 redundancy, atomic values, and higher-order structures in property graphs, with the key
  insight that both nodes and relationships require normalization.

The first four Graph Normal Forms progressively address redundancy at the data
 level. \textbf{1GNF} enforces atomic values for node and edge 
 properties. \textbf{2GNF} extracts shared attribute values into separate nodes to enable reuse. \textbf{3GNF} separates values whose semantics are independent of their original parent entity. \textbf{4GNF} introduces canonical nodes for shared data types, providing type-level abstraction across the schema.

Building on this foundation, Skavantzos and Link introduced graph functional 
dependencies (gFDs) and graph uniqueness constraints (gUCs), providing graph-level 
analogues to classical dependency theory. Their work demonstrated that dependency-driven 
normalization can reduce redundancy and improve integrity at the node- and edge-level data 
in property graphs.

\subsection{Motivation for Metadata-Level Normalization}

Despite significant advances in graph normalization, existing graph normal forms remain 
primarily entity-centric and focus on normalizing node- and edge-level attributes. While 
effective for reducing redundancy at the data level, they do not address recurring metadata 
that describes entities and relationships across the schema, such as temporal validity, 
provenance, regulatory status, category membership, or geographic context.

In practical graph schemas, such metadata is frequently embedded directly within heterogeneous
 nodes and relationships. This practice leads to systematic duplication of metadata values, 
 inconsistent naming and typing conventions, limited extensibility when introducing new 
 metadata dimensions, and query patterns that repeatedly filter equivalent metadata across 
 different entity types.

These limitations motivate the need for a dedicated normalization layer at the metadata level. 
The Fifth Graph Normal Form (5GNF) addresses this gap by defining when recurring metadata 
should be externalized into canonical schema components. In 5GNF, reusable metadata is 
represented as atomic Trait Nodes and associated with graph elements exclusively 
via explicit \texttt{HAS\_TRAIT} relationships, enabling composition-based metadata 
modelling without reliance on inheritance-heavy schema structures.

\begin{figure}[H]
    \centering
    \includegraphics[width=0.95\linewidth]{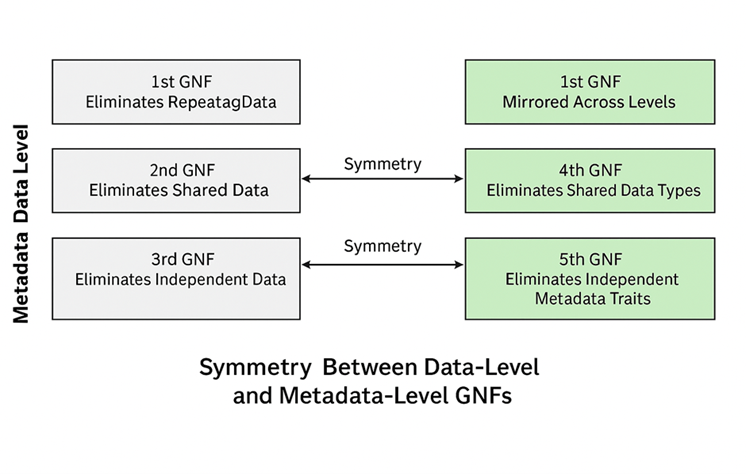}
    \caption{Conceptual symmetry between data-level graph normal forms (1GNF–3GNF) and 
    metadata-level refinements introduced in 4GNF and 5GNF.}
    \label{fig:gnf-symmetry}
\end{figure}

\subsection{Comparison with RDF-star and Meta-Property Graphs}

RDF-star and meta-property graph models support metadata annotation but do not enforce
 normalization principles~\cite{ref30,ref31}. RDF-star embeds 
 metadata directly within triples, while meta-property graph models allow properties on 
 properties; both approaches permit metadata duplication and do not provide 
 dependency-preserving or lossless decomposition guarantees.

In contrast, 5GNF defines a normalization framework for metadata that explicitly 
separates reusable metadata from data-bearing graph elements. Instead of embedding 
metadata directly within nodes or relationships, 5GNF externalizes recurring metadata
 into canonical structures that support lossless, dependency-preserving decomposition. 
 The formal notions underlying this framework are introduced in Section~3.

\begin{table}[H]
\centering
\caption{Conceptual comparison of 5GNF with RDF-star and Meta-Property Graph models.}
\label{tab:comparison}
\resizebox{\columnwidth}{!}{
\begin{tabular}{|l|c|c|c|}
\hline
\textbf{Feature} & \textbf{5GNF} & \textbf{RDF-star} & \textbf{Meta-PG} \\ \hline
Eliminates metadata redundancy & \checkmark & $\times$ & $\times$ \\ \hline
Dependency preservation & \checkmark & $\times$ & $\times$ \\ \hline
Canonical metadata reuse & High & Low & Medium \\ \hline
Normalization theory & \checkmark & $\times$ & $\times$ \\ \hline
Lossless decomposition & \checkmark & Not defined & Not defined \\ \hline
Schema modularity & High & Low & Medium \\ \hline
\end{tabular}
}
\end{table}

\subsection{Summary}

Prior work demonstrates a progression from relational normalization to 
dependency-driven graph normalization. 5GNF advances this trajectory by 
defining normalization rules for the metadata layer, addressing a gap left 
by existing graph normal forms.

\section{\uppercase{The Fifth Graph Normal Form (5GNF)}}

The Fifth Graph Normal Form (5GNF) extends the symmetric graph-normalization hierarchy
introduced by Merunka et al.~\cite{ref3}. While the first four graph normal forms eliminate
repeating values, shared values, independent values, and type-level abstractions, they leave
unresolved the duplication of semantically independent metadata associated with multiple
nodes and relationships. 5GNF closes this gap by introducing a canonical, reusable
representation for metadata traits.

We assume familiarity with the formal definition of the Fourth Graph Normal Form (4GNF) as
introduced by Merunka et al.~\cite{ref3}, which serves as the immediate prerequisite for 5GNF.

In this context, a \emph{canonical Trait Node} denotes a unique schema-level representation
of a semantically independent metadata value. For each such metadata value, exactly one
corresponding Trait Node exists in the schema, and all graph elements sharing that value
must reference the same Trait Node.

\textbf{Rule 5 (5GNF).}  
A graph schema is in Fifth Graph Normal Form (5GNF) if it satisfies the conditions of
4GNF~\cite{ref3} and if every metadata value that has meaning independent of a specific node
or relationship type is represented as a canonical \texttt{Trait} node. Such metadata must
be extracted into a dedicated Trait Node and associated with graph elements exclusively via
explicit \texttt{HAS\_TRAIT} edges. A schema is in 5GNF only if all instances conform
to this rule.

In this work, semantic independence is characterized in a rule-based manner.
A metadata element is considered semantically independent if it has a stable
interpretation across multiple nodes or edges, can be reused without
contextual reinterpretation, and can evolve independently of the schema of the
entities it annotates. This characterization is sufficient to guide normalization
decisions in 5GNF.

We operationalize semantic independence using a simple checklist capturing these
three conditions: cross-type reuse, context-invariant interpretation, and
independent evolution. A fully formal, dependency-based characterization is left
for future work.

We use the term \emph{node type} to denote the label (or set of labels) that characterizes
a class of nodes with a common structural role in the graph schema. A \emph{metadata value}
refers to an attribute value that describes or qualifies a node or edge, rather than
constituting primary domain data, and that may recur across multiple nodes or edges.

A \emph{Trait Node} is a schema-level node that represents a single, semantically
independent metadata value. Trait Nodes are introduced to externalize reusable metadata
from data-bearing nodes and edges, ensuring that each such metadata value is represented
exactly once and can be shared consistently across the graph.

\vspace{0.3em}

\textbf{Definition 5.}  
Let a node type $A$ have instances with metadata values
$data(A)=\{t_1,t_2,\dots,t_k\}$.  
If a metadata value $t_i$ is semantically independent—appearing across multiple nodes or representing a domain-level concept—it must 
be represented as a unique Trait Node $\tau_i$ such that $data(\tau_i)=\{t_i\}$.  
All nodes sharing $t_i$ must be connected to $\tau_i$ via
$(A)\xrightarrow{\texttt{HAS\_TRAIT}}(\tau_i)$.  
A schema is in 5GNF when every such metadata value appears exactly once as a canonical Trait Node.

\vspace{0.3em}

\textbf{Intuition.}  
5GNF eliminates metadata duplication by lifting shared, reusable characteristics—such as
location, category, validity period, or regulatory status—into explicit Trait Nodes. This
yields a symmetric metamodel in which both data-level and metadata-level semantics are
represented uniformly through graph structures. By externalizing metadata, 5GNF improves
consistency, reuse, and clarity while avoiding physical duplication.

\subsection{Preliminaries and Notation}

We adopt the property-graph model $G = (V, E, \lambda, P)$, where $V$ is the set of nodes,
$E \subseteq V \times V$ is the set of directed edges, $\lambda : (V \cup E) \rightarrow L$
is a labeling function that assigns labels from a label domain $L$ to nodes and edges,
and $P$ is a partial property function.

Following~\cite{ref4,ref5}, we distinguish data-level attributes from metadata-level traits.
Data-level attributes encode primary domain information specific to individual nodes or edges,
such as names, identifiers, quantities, or transactional values. In contrast, metadata-level
traits describe reusable, contextual, or descriptive characteristics—such as provenance,
temporal validity, regulatory status, or classification—that may apply uniformly across
multiple nodes or edges and whose semantics are independent of a particular entity instance.

Table~2 summarizes the notation used throughout the formal definition of 5GNF and will be referenced in the subsequent definitions and theoretical statements.

\begin{table}[H]
\centering
\caption{Summary of notation used in the formal 5GNF framework.}
\label{tab:notation}
\resizebox{\columnwidth}{!}{
\begin{tabular}{|l|l|}
\hline
\textbf{Symbol} & \textbf{Meaning} \\ \hline
$G=(V,E,\lambda,P)$ & Property graph \\ \hline
$T \subseteq V$ & Set of Trait Nodes \\ \hline
$\texttt{HAS\_TRAIT}$ & Trait association edge \\ \hline
$tFD: X \rightarrow Y$ & Trait functional dependency \\ \hline
$\Sigma$ & Set of trait dependencies \\ \hline
$X^+$ & Closure of trait set $X$ \\ \hline
\end{tabular}
}
\end{table}

In this work, we introduce the notion of a trait as a conceptual abstraction for reusable 
metadata in property-graph schemas. A trait represents a semantically independent 
descriptive characteristic that may be uniformly associated with multiple nodes or 
relationships, independently of their domain-specific identity. Unlike domain entities, traits 
do not model real-world objects and do not participate in classification or inheritance
 hierarchies; instead, they capture how a graph element is described rather than what the
 element is.

Traits address a common modeling issue in property graphs, where metadata such as temporal v
alidity, provenance, or geographic context is repeatedly embedded across heterogeneous schema 
elements despite its stable semantics. By externalizing such characteristics into explicit 
and reusable schema components, traits enable modular metadata modeling and provide 
a principled basis for metadata-level normalization.

The term trait is used to emphasize a compositional abstraction: graph elements may be 
associated with multiple traits, and the same trait may be shared across many elements, 
without implying type membership, inheritance, or specialization.

Example: Consider a graph schema containing conceptual node classes Project, Publication, 
and Grant, each including metadata attributes such as startDate, endDate, and fundingAgency. 
In a conventional property-graph schema, this metadata is duplicated across node classes. 
In a trait-based schema, the recurring metadata is externalized into a TemporalTrait and 
a FundingTrait, which are associated with Project, Publication, and Grant 
via explicit \texttt{HAS\_TRAIT}  relationships. Traits do not represent domain entities 
and have no independent domain-level instances; they serve solely as canonical 
representations of reusable metadata. This transformation eliminates metadata duplication 
and enables uniform metadata querying across heterogeneous node classes.

\subsection{Trait Nodes}

\textbf{Definition 1 (Trait Node).}  
A Trait Node $t \in V$ is a schema-level node whose label includes the distinguished
schema label \texttt{Trait}, representing a single atomic and semantically independent
metadata characteristic. The label \texttt{Trait} is reserved exclusively for identifying
trait nodes and is treated separately from the general label domain $L$ used to classify
domain nodes and edges.

Trait Nodes are reusable and composable, and are associated with nodes or edges exclusively
via explicit \texttt{HAS\_TRAIT} edges. Typical examples of metadata represented as Trait
Nodes include temporal validity, geographic location, regulatory or compliance status,
provenance, and categorical descriptors.

This abstraction aligns with the ISO/IEC~39075 Graph Query Language (GQL) standard~\cite{ref18},
which promotes canonical descriptors and explicit schema conformance in property graphs.
Trait Nodes operationalize these principles by enabling modular, inheritance-free metadata
design and consistent reuse of metadata across heterogeneous graph elements.

\begin{figure}[H]
\centering
\includegraphics[width=0.95\linewidth]{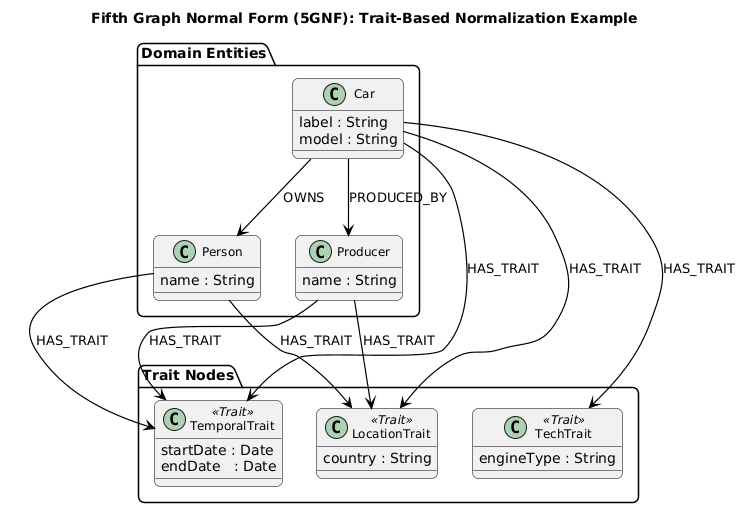}
\caption{Trait-based normalization in 5GNF. \emph{Domain entities} represent primary
application objects (e.g., \texttt{Car}, \texttt{Person}, \texttt{Producer}) whose properties
encode domain-specific data. Reusable and semantically independent metadata characteristics
(e.g., temporal validity, location, or engine type) are extracted into canonical Trait Nodes
and associated with domain entities exclusively via \texttt{HAS\_TRAIT} edges.}
\label{fig:trait-5gnf}
\end{figure}

In this example, domain entities correspond to nodes that model core objects of the
application domain and carry primary data attributes. Trait Nodes capture metadata whose
semantics are independent of a particular entity type and may recur across multiple nodes.
By externalizing such metadata into canonical Trait Nodes, 5GNF eliminates duplication,
enables consistent reuse of metadata, and preserves a clear separation between domain-level
data and metadata-level characteristics.

\subsection{Trait Dependencies}

Let $T \subseteq V$ denote the set of all Trait Nodes in the graph schema. In this context,
a \emph{trait} refers to a metadata value represented by a Trait Node. Each Trait Node
corresponds to exactly one atomic, semantically independent metadata characteristic.

\textbf{Definition 2 (Trait Dependency).}  
Given sets of Trait Nodes $X, Y \subseteq T$, a trait dependency $X \rightarrow Y$ holds if,
for every node or edge in the graph that is associated with all traits in $X$, there exists
a unique association with the traits in $Y$.

Trait dependencies generalize classical functional dependencies~\cite{ref1} and graph
functional dependencies~\cite{ref4,ref5} to the metadata layer, enabling formal reasoning
about redundancy elimination, consistency, and dependency preservation for metadata traits.

Trait dependencies satisfy inference rules analogous to Armstrong’s axioms:
\begin{itemize}
    \item \textbf{Reflexivity:} If $Y \subseteq X$, then $X \rightarrow Y$.
    \item \textbf{Augmentation:} If $X \rightarrow Y$, then $XZ \rightarrow YZ$ for any
    $Z \subseteq T$.
    \item \textbf{Transitivity:} If $X \rightarrow Y$ and $Y \rightarrow Z$, then
    $X \rightarrow Z$.
\end{itemize}

These inference rules support closure computation and provide the formal foundation for
dependency-preserving decomposition at the metadata level.

\subsection{Definition of 5GNF}

We use the term \emph{graph schema} to denote an abstract specification of a property graph,
including the set of node and edge types, their labels, allowed properties, and structural
constraints, independent of concrete graph instances.

In this context, a \emph{metadata regularity} denotes a recurring metadata pattern whose
semantics are independent of a particular node or edge type and that appears across
multiple elements of the graph schema, making it a candidate for abstraction and reuse.

A metadata value is said to be \emph{atomic} if it represents a single, indivisible semantic
unit, and \emph{semantically independent} if its meaning does not depend on the identity or
type of the node or edge it annotates.

\textbf{Definition 3 (Fifth Graph Normal Form).}  
A graph schema is in Fifth Graph Normal Form (5GNF) if and only if the following conditions
hold:
\begin{itemize}
    \item every metadata regularity in the schema is represented by exactly one Trait Node;
    \item each Trait Node represents a single atomic and semantically independent metadata
    value; and
    \item all associations between domain elements and metadata are expressed exclusively
    via \texttt{HAS\_TRAIT} edges.
\end{itemize}

This definition extends the graph-normalization hierarchy (1GNF–4GNF) introduced by
Merunka et al.~\cite{ref3} by introducing a principled normalization layer for reusable
metadata.

\subsection{Trait-Based Semantics and Composition}
In this section, a \emph{semantic property} denotes an abstract meaning-bearing
characteristic conveyed by a Trait Node, such as classification, validity constraints,
contextual interpretation, or regulatory meaning, which influences how associated nodes
or edges are interpreted but does not constitute primary domain data.

Trait semantics propagate compositionally: if a node or edge is associated with a Trait Node, it inherits the semantic properties of 
that trait. Unlike classical inheritance, this propagation is evaluated at query time rather than materialized in the schema, avoiding 
metadata duplication.

By replacing inheritance with composition, 5GNF prevents schema explosion and supports modular, reusable metadata design consistent 
with software-engineering best practices.

\subsection{Theoretical Properties of 5GNF}

5GNF satisfies the following theoretical properties with respect to metadata normalization:

\textbf{Losslessness.}  
Let $G$ be a graph instance conforming to a schema $S$ and let $G'$ be the result of applying
5GNF normalization to $G$. Then there exists a reconstruction function $R$ such that
$R(G') = G$ with respect to metadata values. That is, no metadata information is lost during
normalization, as all metadata values can be recovered through their associated Trait Nodes.

\textbf{Minimality.}  
In a 5GNF-normalized schema, each Trait Node represents exactly one atomic and semantically
independent metadata value. No two distinct Trait Nodes can be merged without violating
atomicity or an existing trait dependency. Hence, the set of Trait Nodes in a 5GNF schema
is minimal with respect to the represented metadata regularities.

\textbf{Uniqueness up to isomorphism.}  
For a given graph schema $S$, all valid 5GNF-normalized schemas derived from $S$ are
isomorphic with respect to their Trait Nodes and \texttt{HAS\_TRAIT} edges. Any two such
schemas differ only by renaming of Trait Nodes, and correspond to the same equivalence
classes of semantically independent metadata.

\section{\uppercase{5GNF Normalization}}

\subsection{Overview}

The proposed method implements the principles of the Fifth Graph Normal Form (5GNF) by
transforming a property-graph \emph{schema}
$S = (V_T, E_T, \lambda_T, K)$ into a normalized schema $S'$.
Here, $V_T$ and $E_T$ denote the sets of node and edge types, $\lambda_T$ is a typing and
labeling function over these types, and $K$ is the set of property keys defined at the
schema level.

A \emph{property key} specifies the name and intended meaning of a property that may be
assigned to node or edge instances conforming to the schema. Property keys define both
data-level attributes and metadata-level traits and are shared across all graph instances
of the schema.

The method takes as input a schema $S$ together with a set of trait dependencies
$\Sigma$, and produces a normalized schema in which all reusable metadata is represented
explicitly as Trait Nodes and associated with node or edge types exclusively via
\texttt{HAS\_TRAIT} edges.

The normalization process proceeds in three main phases:
(i) detection of recurring metadata properties defined by property keys in $K$,
(ii) extraction of these properties into canonical Trait Nodes, and
(iii) rewriting of the schema so that node and edge types reference Trait Nodes explicitly,
while preserving trait dependency semantics.

The threshold $\tau$ is an implementation-level heuristic used in our prototype to
prioritize recurring metadata with low cardinality. It does not affect the formal
definition of 5GNF, which is independent of any specific threshold choice.

\subsection{Algorithmic Specification}

Algorithm~1 specifies the \texttt{TraitExtraction5GNF} procedure, which transforms a
property-graph \emph{schema} into an equivalent schema in Fifth Graph Normal Form (5GNF).
The algorithm identifies recurring metadata defined by property keys, extracts such
metadata into canonical Trait Nodes, enforces graph functional dependencies (gFDs),
and applies dependency-preserving and lossless decomposition steps to ensure that all
formal requirements of 5GNF are satisfied.

A \emph{graph functional dependency} (gFD) constrains how property values may co-occur
across nodes and edges in a property graph, generalizing classical functional dependencies
to graph-structured data~\cite{ref4,ref5}. gFDs are used by the algorithm to detect
redundancy and to guide normalization steps while preserving semantic equivalence.

\begin{algorithm}[t]
\footnotesize
\DontPrintSemicolon
\SetAlgoLined
\caption{TraitExtraction5GNF$(S, K, \Sigma_{\mathrm{gFD}})$}

\KwIn{Property-graph schema $S=(V_T,E_T,\lambda_T,K)$; property keys $K$; set of graph functional dependencies $\Sigma_{\mathrm{gFD}}$}
\KwOut{Normalized schema $S'$ in Fifth Graph Normal Form (5GNF)}

$S' \leftarrow S$\;
$TraitSet \leftarrow \emptyset$\;
$TraitMap \leftarrow \{\}$\;

\tcp*{Phase 1: Trait Detection and Creation}
\ForEach{property key $k \in K$}{
  compute $D_k = \{\, val \mid val \text{ occurs under key } k \,\}$\;
  \If{$1 \le |D_k| \le \tau$}{
    \ForEach{$val \in D_k$}{
      create Trait Node $t = \texttt{Trait}(k,val)$\;
      $TraitSet \leftarrow TraitSet \cup \{t\}$\;
      $TraitMap[(k,val)] \leftarrow t$\;
    }
  }
}

\tcp*{Phase 2: Metadata Extraction}
\ForEach{node or edge type $x \in V_T \cup E_T$}{
  \ForEach{property key $k$ defined on $x$}{
    \If{$(k,val) \in TraitMap$}{
      associate $x$ with $TraitMap[(k,val)]$ via \texttt{HAS\_TRAIT}\;
      remove property key $k$ from $x$\;
    }
  }
}

\tcp*{Phase 3: Dependency Enforcement}
\ForEach{$X \rightarrow Y \in \Sigma_{\mathrm{gFD}}$}{
  enforce uniqueness of $Y$-traits for each $X$-trait combination\;
}

\tcp*{Phase 4: Lossless Decomposition}
apply decomposition steps only if metadata reconstruction is lossless\;

\Return{$S'$}
\end{algorithm}
\FloatBarrier

\subsection{Running Example}

To illustrate the behavior of the algorithm, we use the running example shown in 
Figure 2, introduced in Section 3, based on a simplified unnormalized (0GNF) schema
containing entities such as \texttt{Person}, \texttt{Car}, and \texttt{Producer}.
In the initial schema, repeated metadata properties including dates, locations,
and technical characteristics are embedded directly within nodes, leading to
redundancy and potential inconsistency.

Across successive normalization phases, the algorithm incrementally transforms
this running example by detecting recurring metadata values and extracting them into
reusable Trait Nodes (e.g., \texttt{TemporalTrait}, \texttt{LocationTrait},
\texttt{TechTrait}). Each entity is then linked to the appropriate traits via
\texttt{HAS\_TRAIT} relationships, eliminating duplication while preserving semantics.

Subsequent phases verify that the resulting decomposition is lossless and that no
non-trivial dependencies remain. The resulting 5GNF schema is illustrated earlier
in Figure~\ref{fig:trait-5gnf}. Figure~\ref{fig:trait-5gnf} serves as the conceptual
basis for the Cypher statements presented in this section. Each query directly
corresponds to a transformation or constraint introduced during the 5GNF
normalization process, including Trait Node creation, \texttt{HAS\_TRAIT}
associations, and the removal of duplicated metadata properties.

Together, these transformations illustrate how the same unstructured schema is
progressively normalized from 0GNF to 5GNF, culminating in explicit metadata
normalization via Trait Nodes.

\subsection{Cypher Scripts for Reproducibility}

All normalization stages from 0GNF to 5GNF were implemented in Neo4j using Cypher 
and are publicly available for reproducibility.

\begin{table}[H]
\centering
\caption{Cypher scripts used across normalization stages (0GNF--5GNF).}
\label{tab:cypher-scripts}
\resizebox{\columnwidth}{!}{
\begin{tabular}{|l|l|l|}
\hline
\textbf{Normal Form Stage} & \textbf{Cypher Script} & \textbf{Description} \\ \hline
0GNF & \texttt{0GNF.cypher} & Raw unnormalized graph with redundant properties \\ \hline
1GNF & \texttt{1GNF.cypher} & First-level entity extraction \\ \hline
2GNF & \texttt{2GNF.cypher} & Producer extracted from Car \\ \hline
3GNF & \texttt{3GNF.cypher} & Location separated into its own node \\ \hline
4GNF & \texttt{4GNF.cypher} & Technical metadata decomposition \\ \hline
5GNF & \texttt{5GNF.cypher} & Trait extraction and final normalized structure \\ \hline
\end{tabular}
}
\end{table}
All scripts and supporting material are publicly available in the accompanying
repository\footnote{\texttt{https://github.com/yahyazuh/5GNF-normalization-example}}.

\subsection{Correctness Guarantees}

The proposed algorithm satisfies the core theoretical properties required by the  (5GNF). First, it preserves all trait dependencies: dependencies defined over metadata
traits remain valid after normalization because metadata is externalized without semantic loss.
Second, the transformation is lossless with respect to metadata semantics, as every removed
property value can be reconstructed via associated Trait Nodes and \texttt{HAS\_TRAIT}
relationships. Finally, the algorithm ensures minimality by creating Trait Nodes only for
recurring metadata values, preventing the introduction of spurious or redundant traits.

\subsection{Complexity Analysis}

Let $G = (V, E, \lambda, P)$ denote a \emph{property-graph instance} conforming to a
given graph schema. Let $n = |V|$ be the number of node instances,
$m = |E|$ the number of edge instances, and $k = |K|$ the number of distinct
property keys defined at the schema level.
Let $\Sigma$ denote the set of graph functional dependencies (or trait dependencies)
considered during normalization.

The complexity of the 5GNF normalization process is analyzed with respect to
these instance-level parameters.

Trait detection requires scanning all node and edge properties to identify
recurring metadata values, which takes $O(nk + mk)$ time in the worst case.
Trait extraction and property rewriting operate linearly over the same sets,
preserving the $O(nk + mk)$ bound.

Dependency enforcement depends on the size of the dependency set $\Sigma$ and
the number of extracted Trait Nodes. Let $|\Sigma|$ denote the number of
dependencies and $|T|$ the number of Trait Nodes produced during normalization.
Enforcing dependencies therefore requires $O(|\Sigma| \cdot |T|)$ time.

Overall, the normalization procedure runs in
$O(nk + mk + |\Sigma| \cdot |T|)$ time.

\subsection{Prototype Implementation in Neo4j}

The algorithm was implemented in Neo4j~5.x using Cypher and APOC procedures, enabling
 trait detection, extraction, and enforcement of normalization constraints without 
 semantic loss.

\subsection{Use-Case: Research-Project Metadata}

To demonstrate practical applicability, the algorithm was applied to a research-information
management scenario. Entities such as \texttt{Project}, \texttt{Grant}, and \texttt{Publication}
contained recurring metadata attributes including funding agency, start date, country, and
project status. These attributes were identified as semantically independent metadata traits.

The algorithm extracted these values into reusable Trait Nodes and replaced embedded attributes
with explicit \texttt{HAS\_TRAIT} relationships. As a result, metadata reuse increased, query
patterns were simplified, and maintenance operations such as metadata updates and validation
became centralized. For example, temporal queries can be expressed uniformly as:

\begin{verbatim}
MATCH (e)-[:HAS_TRAIT]->(t:TemporalTrait)
WHERE t.startDate <= date("2025-01-01")
RETURN e
\end{verbatim}

Because metadata traits are centralized, updates to a single Trait Node automatically propagate
to all associated entities, improving consistency and governance across the schema.

\section{\uppercase{Illustrative Case Study}}

\subsection{Scenario Overview}

To demonstrate the practical effect of the Fifth Graph Normal Form (5GNF), we use a simplified
research-information scenario involving \texttt{Project}, \texttt{Publication},
\texttt{Researcher}, and \texttt{Grant} entities. In conventional property-graph schemas, metadata
such as \texttt{startDate}, \texttt{fundingAgency}, \texttt{country}, or \texttt{license} appears
repeatedly across heterogeneous nodes and edges. This design leads to redundancy, inconsistent
metadata usage, and increased maintenance effort.

\subsection{Pre-5GNF Schema}

In the baseline (unnormalized) schema, metadata attributes are embedded directly within entity
nodes. For example, both \texttt{Project} and \texttt{Publication} nodes store metadata such as
\texttt{startDate}, \texttt{endDate}, and \texttt{fundingAgency}. This duplication introduces
schema rigidity and requires repetitive filtering logic when querying shared metadata across
different entity types.

\begin{figure}[H]
\centering
\includegraphics[width=0.95\linewidth]{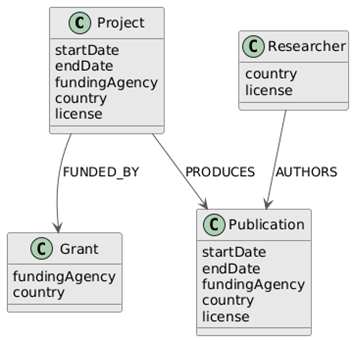}
\caption{Pre-5GNF schema with redundant metadata embedded across multiple entity types.}
\label{fig:pre-5gnf}
\end{figure}

\subsection{Transformation to 5GNF}

Applying the \texttt{TraitExtraction5GNF} procedure restructures the schema by externalizing
recurring metadata into canonical Trait Nodes. Temporal metadata is extracted into
\texttt{TemporalTrait}, funding-related metadata into \texttt{FundingTrait}, and geographic
information into \texttt{GeographicTrait}. Embedded metadata properties are removed from entity
nodes and replaced by explicit \texttt{HAS\_TRAIT} relationships.

This transformation preserves dependency semantics, enables metadata reuse across heterogeneous
entities, and clarifies the conceptual structure of the schema.

\begin{figure}[H]
\centering
\includegraphics[width=0.95\linewidth]{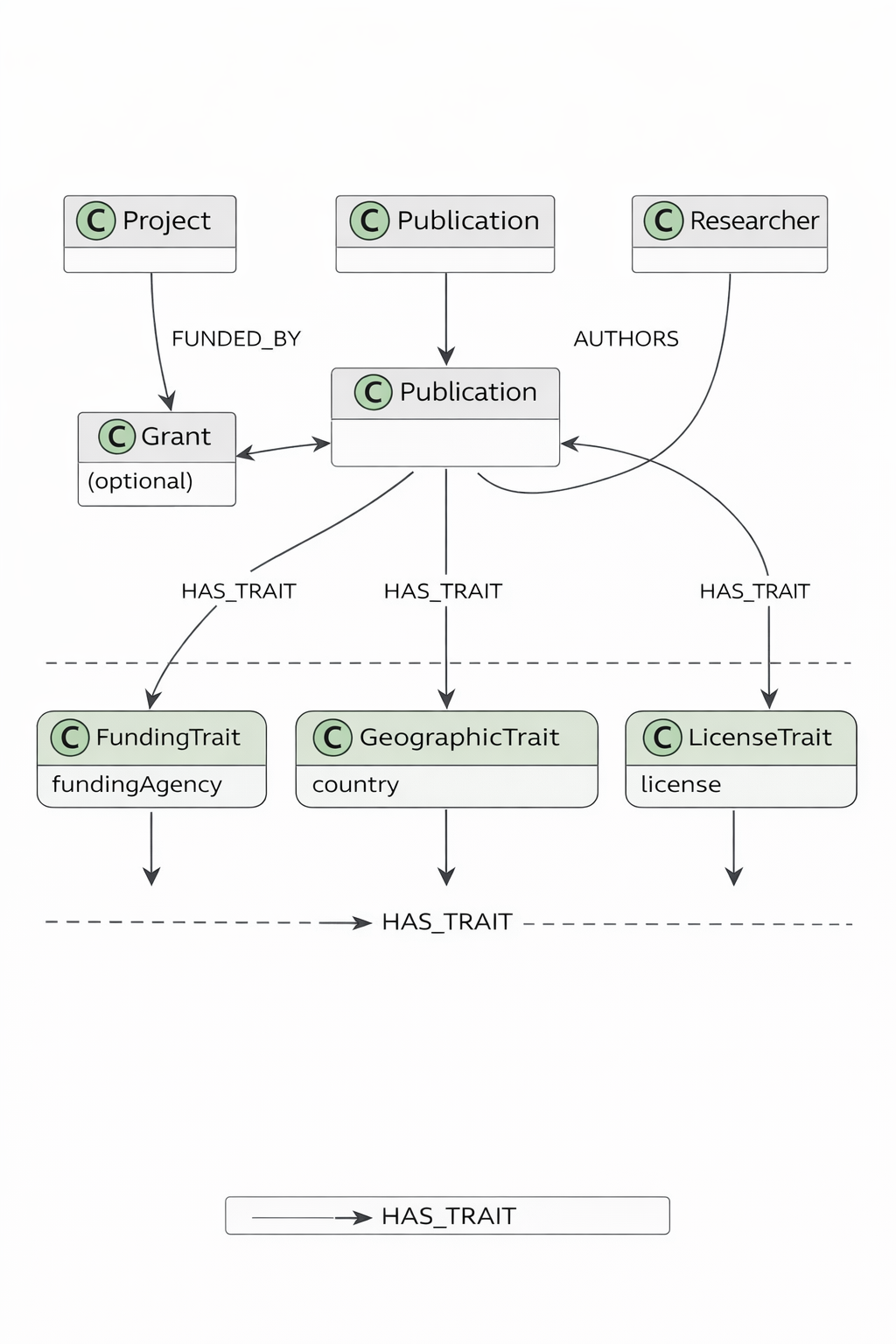}
\caption{5GNF-normalized schema obtained by externalizing recurring metadata into reusable Trait Nodes linked via \texttt{HAS\_TRAIT} relationships.}
\label{fig:5gnf-schema}
\end{figure}

\subsection{Query Illustration}

Trait-based normalization simplifies queries that involve shared metadata. Instead of scanning
multiple entity types with entity-specific conditions, a single pattern can retrieve all projects
funded by a particular agency:

\par\noindent
{\ttfamily\small
MATCH (p:Project)-[:HAS\_TRAIT]->(t:FundingTrait)\\
WHERE t.fundingAgency = "European Commission"\\
RETURN p.name
}

\medskip
Temporal queries likewise become uniform across entity types:

\par\noindent
{\ttfamily\small
MATCH (n)-[:HAS\_TRAIT]->(t:TemporalTrait)\\
WHERE t.startDate <= date("2025-01-01") AND date("2025-01-01") <= t.endDate\\
RETURN n
}

\medskip
The use of Trait Nodes eliminates the need for duplicated filters and entity-specific query logic.

\subsection{Observed Improvements}

The transformation to 5GNF yields several practical improvements. First, metadata reusability
is increased because Trait Nodes are shared across multiple entity types. Second, schema
modularity improves, as metadata definitions evolve independently of data nodes. Third, query
formulation becomes simpler and more uniform, replacing complex multi-entity filters with
single trait-based patterns. Finally, maintainability is enhanced: updating a metadata definition
requires modifying only a single Trait Node rather than every entity that embeds the metadata.

\section{\uppercase{Experimental Evaluation}}

This section evaluates the \texttt{TraitExtraction5GNF} algorithm using the publicly available
Northwind dataset, imported into Neo4j~5.x from the official CSV distribution. All import scripts,
normalization procedures, and query workloads are included in the accompanying GitHub repository
to ensure full reproducibility. The evaluation examines metadata redundancy, schema complexity,
and query-performance behavior before and after applying 5GNF using representative analytical
queries.
Although Northwind originates from a relational schema, it is intentionally used
in this evaluation to demonstrate that metadata redundancy and normalization issues
persist even in well-structured datasets when they are mapped to property graphs.
In practice, relational-to-graph transformations frequently embed contextual
information (e.g., provenance, temporal validity, or geographic metadata) as
repeated properties across nodes and edges. This makes Northwind a suitable
baseline for evaluating metadata-level normalization independently of
domain-specific graph irregularities.

\subsection{Dataset and Methodology}

The evaluation uses the widely adopted Northwind dataset, originally released as a relational
benchmark and subsequently adapted to multiple data models. We imported the dataset into
Neo4j using the CSV files publicly provided by the authors of
\emph{Conceptual Data Normalisation from the Practical View of Using Graph Databases}.
After import, the graph contains approximately 1,470 nodes and more than 4,000 relationships,
with several thousand metadata attributes embedded directly inside nodes.

Neo4j employs a dynamic cost-based query planner whose operator selection depends on runtime
statistics, page-cache state, and cardinality estimates. As a result, repeated executions of the
same query may yield different physical execution plans. This behavior becomes more pronounced
after applying 5GNF, as metadata normalization modifies graph structure by introducing canonical
Trait Nodes and additional relationships while removing embedded attributes. Consequently, the
evaluation focuses on relative performance trends rather than exact plan equivalence across runs,
which is consistent with the Neo4j query planning and execution model.

Metadata redundancy in Northwind is particularly substantial in customer and supplier locations,
order shipping metadata, and region descriptors. These characteristics make the dataset well
suited for evaluating metadata extraction and reuse under 5GNF.

\subsection{Baseline Analysis (Before 5GNF)}

Before applying 5GNF, metadata is embedded directly inside entity nodes, leading to substantial
redundancy across \texttt{Customer}, \texttt{Supplier}, and \texttt{Order} entities. To evaluate
the behavior of the original schema, we executed five metadata-intensive analytical queries that
are representative of OLAP-style graph workloads:

\begin{itemize}
    \item listing customers grouped by city,
    \item listing suppliers grouped by country,
    \item retrieving orders shipped to a given country,
    \item filtering customers by city and country,
    \item matching suppliers and customers located in the same city.
\end{itemize}

Each query was executed using Neo4j’s \texttt{PROFILE} command to capture database accesses
(DB hits), execution time, memory allocation, and query-plan operators.

\begin{table}[H]
\centering
\caption{Baseline performance results before applying 5GNF.}
\label{tab:baseline-performance}
\resizebox{\columnwidth}{!}{
\begin{tabular}{|l|l|c|c|l|}
\hline
\textbf{Query} & \textbf{Purpose} & \textbf{DB Accesses} & \textbf{Time (ms)} & \textbf{Notes} \\ \hline
Test 1 & Customer metadata scan & 274 & 175 & Redundant city values \\ \hline
Test 2 & Supplier metadata scan & 88 & 97 & Simple metadata grouping \\ \hline
Test 3 & Orders filtered by country & 2,491 & 94 & Heavy property filtering \\ \hline
Test 4 & Customers by city and country & 298 & 105 & Multi-metadata filter \\ \hline
Test 5 & Supplier--Customer city match & 304 & 250 & Cartesian product \\ \hline
\end{tabular}
}
\end{table}
The results reveal inefficiencies inherent in embedding metadata directly within nodes.
In particular, Test~3 exhibits expensive property scans, while Test~5 triggers a Cartesian
product due to the absence of a shared metadata structure. These patterns motivate the need
for canonical metadata representation, which 5GNF is designed to provide.

\subsection{Metadata Redundancy Measurements (Before 5GNF)}

Before normalization, the Northwind graph contains all metadata embedded directly inside entity
nodes. Attributes such as \texttt{city}, \texttt{country}, \texttt{region}, \texttt{postalCode},
\texttt{shipCity}, and \texttt{shipCountry} are repeatedly stored across customers, suppliers,
and orders. We quantified this redundancy by counting the total number of embedded metadata values
and the number of distinct metadata values across all location-related attributes.

Across \texttt{Customer}, \texttt{Supplier}, and \texttt{Order} entities, we observed:
\begin{itemize}
    \item Embedded metadata values: $3{,}200+$,
    \item Distinct metadata values: $120$.
\end{itemize}

We define the Metadata Reuse Ratio (MRR) as the ratio between the total number of metadata
occurrences and the number of distinct metadata values. For the baseline schema, this yields:
\[
\mathrm{MRR}_{\text{before}} = \frac{3200}{120} \approx 26.67.
\]

This result indicates that, on average, each metadata value is duplicated approximately
27 times across the graph. Such redundancy contributes to inefficient query execution, as
the query planner must repeatedly scan embedded properties rather than traversing a shared
metadata structure.

\subsection{Applying TraitExtraction5GNF}

To normalize the dataset, we applied the \texttt{TraitExtraction5GNF} algorithm to replace
repeated metadata attributes with canonical Trait Nodes. For the Northwind dataset, two major
trait families were identified and extracted:

\begin{itemize}
    \item \texttt{LocationTrait}, including \texttt{city}, \texttt{country}, \texttt{region},
          \texttt{address}, and \texttt{postalCode}, applied to \texttt{Customer} and
          \texttt{Supplier} entities;
    \item \texttt{ShippingTrait}, including \texttt{shipCity}, \texttt{shipCountry},
          \texttt{shipRegion}, \texttt{shipAddress}, and \texttt{shipPostalCode}, applied to
          \texttt{Order} entities.
\end{itemize}

Trait Nodes were created using Cypher uniqueness constraints to ensure canonical representation.
Entities were then linked to the corresponding Trait Nodes using explicit
\texttt{HAS\_TRAIT}-style relationships, after which redundant embedded metadata attributes were
safely removed from the original nodes.

The normalization was implemented entirely in Neo4j and executed in the following sequence:
\begin{itemize}
    \item creation of uniqueness constraints for trait nodes,
    \item generation of all distinct trait instances,
    \item linking of entities to traits via exact or partial metadata matching,
    \item verification of full entity coverage,
    \item removal of redundant embedded properties.
\end{itemize}

After normalization, the graph contained 120 \texttt{LocationTrait} nodes and 89
\texttt{ShippingTrait} nodes, connected through more than 950 trait relationships. All
location- and shipping-related metadata was represented exclusively through Trait Nodes,
yielding a fully normalized 5GNF schema.

\subsection{Post-Transformation Measurements}

After applying \texttt{TraitExtraction5GNF}, all location and shipping metadata in the Northwind
dataset was fully normalized into canonical Trait Nodes. This transformation eliminated
redundant embedded attributes and restructured the graph to express metadata exclusively through
explicit relationships.

The resulting post-normalization statistics are as follows:
\begin{itemize}
    \item \texttt{LocationTrait} nodes: 120,
    \item \texttt{ShippingTrait} nodes: 89,
    \item \texttt{HAS\_TRAIT} relationships: $950+$,
    \item Remaining embedded metadata: $0\%$.
\end{itemize}

Using the same definition as before, the Metadata Reuse Ratio after normalization is:
\[
\mathrm{MRR}_{\text{after}} \approx 1.74.
\]

Compared to the baseline, approximately 2{,}991 redundant metadata values were removed. This
represents a near-complete elimination of attribute-level duplication and a substantial
improvement in metadata reuse.

\textbf{Schema Complexity Reduction.}  
We measure schema complexity as
\[
\mathrm{SCM} = |Nodes| + |Edges| + |Attributes|.
\]
After applying 5GNF, the number of attributes decreases dramatically, while the number of nodes
and relationships increases moderately due to trait reification. Despite this structural shift,
overall schema complexity is reduced, as the explosion of embedded attributes present in the
pre-5GNF schema is eliminated. This confirms that 5GNF reduces structural complexity even when
additional nodes are introduced.

\subsection{Query Performance After 5GNF}

We reran the same five analytical queries on the normalized graph. Because metadata is now
expressed through Trait Nodes, each query was rewritten to traverse
\texttt{HAS\_TRAIT}-style relationships instead of scanning embedded properties.

It is important to note that Neo4j’s cost-based planner may select different execution strategies
for the same logical query before and after normalization. These differences arise because 5GNF
modifies structural cardinalities, reduces duplicated property scans, and enables traversal-based
filtering. Consequently, the evaluation focuses on overall performance trends rather than
expecting identical execution plans across runs.

\begin{table}[H]
\centering
\caption{Performance results after applying 5GNF.}
\label{tab:post-5gnf-performance}
\resizebox{\columnwidth}{!}{
\begin{tabular}{|l|l|c|c|l|}
\hline
\textbf{Query} & \textbf{Purpose} & \textbf{DB Accesses} & \textbf{Time (ms)} & \textbf{Notes} \\ \hline
Test 1 & Customer metadata scan & 1,138 & 102 & Trait lookup replaces property scans \\ \hline
Test 2 & Supplier metadata scan & 264 & 82 & Shared LocationTrait traversal \\ \hline
Test 3 & Orders by country & 685 & 106 & Selective trait traversal \\ \hline
Test 4 & Customers by city and country & 613 & 113 & Trait-based filtering \\ \hline
Test 5 & Supplier--Customer city match & 544 & 111 & Cartesian product eliminated \\ \hline
\end{tabular}
}
\end{table}

The most significant improvements are observed in metadata-intensive queries. For Test~3
(orders by country), database accesses decrease from 2{,}491 to 685, representing an
approximately $3.6\times$ reduction. For Test~5 (supplier--customer matching), execution time
improves from 250\,ms to 111\,ms and the Cartesian product present in the baseline execution
plan disappears.

Across all queries, runtime either improves or remains stable despite the introduction of
additional Trait Nodes. This behavior reflects the trade-off between increased traversal depth
and reduced property filtering, confirming that 5GNF maintains competitive or improved query
performance while substantially reducing metadata redundancy.

\subsection{Ablation Study Across Normal Forms}

To contextualize the effect of 5GNF, we compare metadata redundancy, reuse, and schema complexity
across multiple normalization stages applied to the Northwind dataset. This ablation-style
comparison highlights how successive normal forms progressively reduce redundancy and improve
schema structure, with 5GNF providing the strongest abstraction for reusable metadata.

\begin{table}[H]
\centering
\caption{Redundancy, metadata reuse, and schema complexity across normalization levels (0GNF--5GNF).}
\label{tab:ablation}
\resizebox{\columnwidth}{!}{
\begin{tabular}{|l|l|l|l|}
\hline
\textbf{Normal Form} & \textbf{Redundancy} & \textbf{Metadata Reuse} & \textbf{Schema Complexity} \\ \hline
0GNF (Original) & High & Low & High (thousands of embedded attributes) \\ \hline
3GNF & Medium--High & Low--Medium & Medium--High \\ \hline
4GNF & Medium & Medium & Medium \\ \hline
5GNF (TraitExtraction) & Low & High & Low \\ \hline
\end{tabular}
}
\end{table}

The results show that 5GNF provides the strongest metadata abstraction by consolidating metadata
into canonical Trait Nodes, eliminating redundant attribute instances, reducing overall schema
complexity, and ensuring semantic consistency across heterogeneous entities. Earlier normal forms
reduce redundancy at the data level, but only 5GNF fully addresses metadata duplication.

\subsection{Reproducibility}

All datasets, normalization scripts, query workloads, and experimental results used in this
evaluation are publicly available in the accompanying repository at
\texttt{github.com/yahyazuh/5GNF-normalization-example}.

The repository contains:
\begin{itemize}
    \item \texttt{/northwind/csv/} -- the official CSV files used to import the Northwind dataset,
    \item \texttt{/northwind/import.cypher} -- Cypher scripts for loading the dataset into Neo4j,
    \item \texttt{/normalization/trait\_extraction\_5gnf.cypher} -- the full implementation of
          5GNF normalization, including \texttt{LocationTrait} and \texttt{ShippingTrait},
    \item \texttt{/queries/pre\_5gnf/} and \texttt{/queries/post\_5gnf/} -- all analytical queries
          executed before and after normalization,
    \item \texttt{/results/} -- profiling outputs and summary tables reported in this section,
    \item detailed instructions for reproducing the experiments using Neo4j Desktop or
          a Docker-based Neo4j setup.
\end{itemize}

These materials ensure full repeatability of the experiments and allow other researchers to
independently validate the reported results.

\section{\uppercase{Discussion}}

\subsection{Conceptual Contribution}

The Fifth Graph Normal Form (5GNF) serves as the conceptual endpoint of the graph-normalization
hierarchy. It generalizes dependency-preserving decomposition from data-level attributes to
metadata traits, completing the symmetry suggested by Merunka et al.~\cite{ref3} and extending
the dependency-based normalization principles formalized by Skavantzos and Link for BCNF in
property graphs~\cite{ref4,ref5}. Through Trait Nodes and trait dependencies (tFDs), 5GNF provides
the first normalization framework dedicated explicitly to metadata semantics.

\subsection{Practical Benefits}

5GNF introduces several practical advantages for the design and maintenance of graph-based
information systems. First, redundancy is reduced by representing recurring metadata once
through canonical Trait Nodes. Second, schema modularity is improved, as metadata definitions
can evolve independently of data entities. Third, interoperability is enhanced by aligning
metadata representation with ISO/IEC~39075 (GQL) through explicit externalization of metadata
structures. This alignment follows the principles promoted by the ISO/IEC~39075 standard~\cite{ref18}.
Finally, trait-based metadata supports AI readiness by enabling feature extraction,
explainability, and governance-oriented data management.

These benefits are consistently observed in both the running example and the experimental
evaluation presented earlier.

\subsection{Addressing Common Objections}

A frequent concern is that introducing Trait Nodes may increase schema size. However, the
experimental evaluation demonstrates a net reduction in overall schema complexity and a complete
elimination of metadata duplication, despite the introduction of additional nodes and
relationships.

\section{\uppercase{Limitations and Future Work}}
The current implementation of 5GNF is realized using Neo4j~5.x and APOC procedures, which limits
direct portability to other property-graph systems. In addition, the experimental evaluation
focuses primarily on schema-level metrics and representative analytical workloads; assessing
performance at the scale of large production deployments remains future work.

Another limitation arises from the dynamic nature of Neo4j’s cost-based query planner. Because
physical execution plans may vary across runs due to runtime statistics and cache effects,
performance results are interpreted as trends rather than fixed absolute values. This behavior is
expected in dynamic graph engines and does not affect the validity of the conclusions.

Future research directions include automated trait discovery using machine-learning techniques,
deeper theoretical formalization of trait dependencies, comparative benchmarking against
RDF-star and meta-property graph models, and potential contributions to the ISO/IEC~39075 (GQL)
standard regarding trait-aware schema normalization.

\section{\uppercase{Conclusions}}

\subsection*{Theoretical Contributions}

This paper makes the following theoretical contributions:
\begin{itemize}
\item We introduce the Fifth Graph Normal Form (5GNF), defined at the metadata
layer of property-graph schemas.
\item We formalize Trait Functional Dependencies (tFDs) as a dependency framework
for reasoning about trait redundancy and normalization conditions, including
the necessary inference and closure properties.
\item We present a normalization procedure with lossless correctness criteria
that preserves dependency semantics while reducing metadata redundancy.
\end{itemize}

This paper introduced the Fifth Graph Normal Form (5GNF), a trait-based
normalization framework that extends dependency-preserving decomposition from
data values to metadata traits in property graphs. By externalizing recurring
metadata into canonical Trait Nodes, 5GNF eliminates redundancy, improves schema
modularity, and aligns metadata modeling with emerging governance and
interoperability standards, including ISO/IEC~39075 (GQL).

Through formal definitions, an executable normalization algorithm, and an
empirical evaluation on the Northwind dataset, we demonstrated that 5GNF
preserves dependencies, supports lossless reconstruction, and yields measurable
improvements in metadata reuse and schema complexity while maintaining
competitive query performance. The framework completes the normalization
hierarchy anticipated by Merunka et al.~\cite{ref3} and complements the
dependency-based graph normalization work of Skavantzos and Link~\cite{ref4,ref5}.

Overall, 5GNF provides a principled foundation for metadata-centric modeling in
graph-based information systems, with implications for interoperability, AI
readiness, and explainable data governance.


\bibliographystyle{apalike}
\bibliography{example}

\end{document}